# Windshield Integration of Thermal and Color Fusion for Automatic Emergency Braking in Low Visibility Conditions

Gabriel Jobert[1], Guillaume Delubac[1], Jessy Matias[1], Quentin Noir[1], Xavier Brenière[1], Pauline Girard[2], Tatiana Severin-Fabiani[3], Sebastien Tinnes[1]

1: LYNRED, 364 route de Valence, 38113 Veurey-Voroize, France
2: Saint-Gobain Research Compiègne, 1 Rue de Montluçon, 60150 Thourotte, France
3: Saint-Gobain Research Paris, 39 Quai Lucien Lefranc, 93300 Aubervilliers, France

**Abstract**: The new NHSTA regulations require Automatic Emergency Braking (AEB) systems to operate at night, to protect pedestrians in the deadliest conditions. We propose thermal imaging as a new sensor to complement the AEB sensor suite, alongside the visible front camera and RADAR. This paper explores the benefits of visible-thermal fusion, proposes a windshield integration for such a system, and evaluates the minimum performance requirements for a thermal camera compliant with the NHSTA standards, based on a field study of pedestrian detection range.

## 1. Introduction

According to the European Road Safety Observatory [1], more than half of the road fatalities are vulnerable road users, and 23% are pedestrians. Similar statistics are reported by the National Highway Traffic Safety Administration (NHTSA) [2] in USA's roads. Moreover, about 70% of the total share of pedestrian fatalities in the USA and Europe occurs during adverse conditions: at night, and/or with poor weather, and on poorly lighted rural roads.

Current Automatic Emergency Braking (AEB) systems rely on a combination for the visible front camera and a RADAR, and are able to cope with most scenarios as long as the ambient lighting condition is good (>500lux). However, it fails at night where most pedestrian fatalities occur. The NHTSA adopted in May 2024 a new regulation (effective in 2029) that extends the AEB operational design domain to low light conditions (<0.2 lux) [3]. For instance, a car to pedestrian crash must be avoided for a car at a speed of 60km/h (37.5mph).

Thermal infrared imaging, using an uncooled microbolometer array, is already integrated in Advanced Driver Assistance Systems (ADAS) to facilitate nighttime driving. We will detail why thermal imaging, fused with the visible, can be used to extend the operational design domain of an AEB system to nighttime conditions, and more generally reduced visibility conditions (rain, fog, glare, etc.). We will then demonstrate the integration of a thermal camera into the windshield, using a process co-developed with Saint-Gobain Sekurit. Finally, we will define the minimum specifications for both the thermal sensor and the objective lens to be compliant with the NHSTA requirements, based on field trials of AI-based pedestrian detection range. This will allows us to scale down the AEB thermal camera, thanks to the latest advances in manufacturing technologies of small sized sensors and lenses.

## 2. Visible-Thermal fusion in ADAS

**2.1. Specificities of visible and thermal imaging:** A typical visible RGB camera operates in the visible range (0.4-0.7µm), and captures a color image of reflected light by the scene, from an external light source, such as sunlight, street lamps, or the vehicle's own headlights. While looking natural to the human eye, the scene can be highly complex to interpret in the sense of computer vision, that relies on texture and color-albedo contrast.

Thermal imaging technologies originate from defense and security applications (such as night-vision googles, thermal weapon sights or drones), and allow for detecting threats during both day and night. This application led to the development and continuous improvements of uncooled micro-bolometers arrays, while lowering the SWaP (Size-Weight-and-Power). Such a thermal camera operates in the Long-Wave-InfraRed (LWIR) range (8 – 14 µm) and captures a monochrome image directly emitted by the scene itself, without illumination. The radiated heat depends on the target emissivity and temperature, according to Planck's law.

This fundamental difference between visible and thermal images (reflective vs. emissive) makes the first highly dependent on the lighting conditions, and not the other. In an unlit environment, the visible camera can only see as far as the vehicle's headlights can reach. The intensity of the reflected light decreases according to the inverse of the distance to the power of four (round-trip of light). The detection range of the thermal camera mainly depends on its spatial resolution. Lighting comes from the scene: a human is an infrared light bulb!

In Figure 1, we see an image comparison of a nighttime driving scene. A distant pedestrian (see, digital-zoom view), beyond the range of the headlights, is only detected by the thermal camera.

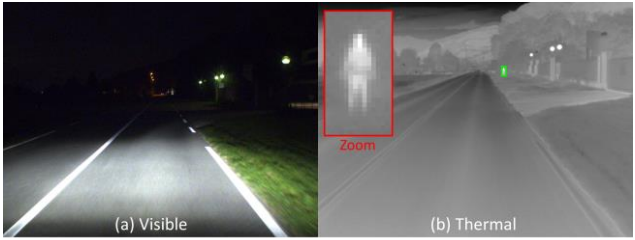

*Figure 1: Image comparaison of a night-time scene: (a) visible and (b) thermal.*

*Image processing and inference:* In this document, the 14-bits raw thermal images are simply processed using Lynred's shutter-less non-uniformity correction, and then rescaled to 8-bits images. In order to limit the complexity of the study, we do not apply additional post-processing to the images (such as denoising, local contrast enhancements, super-resolution etc.).

We detect pedestrians, using a CNN (convolutional neural network) on either 8-bit grayscale thermal images or 8-bit RGB color images. The model is based on a ResNet50 [4], fine-tuned for each modalities on only 5.7k labelled visible/thermal pairs from Lynred's automotive fusion dataset. Note that, the green bounding box in Figure 1 locates the pedestrian detected by the network.

The detection results presented in this document depend on the training dataset, the choice of image processing parameters, and the neural network architecture. Performances variance due to thermal image processing parameters and detection model are found in reference [5].

**2.2. Fusing complementary modalities:** The visible camera is very efficient with good lighting conditions, and provides high-resolution color images. Thermal is interesting as a complementary modality that reveals new physical properties of the scene, improving reliability of AEB systems in all conditions [6].

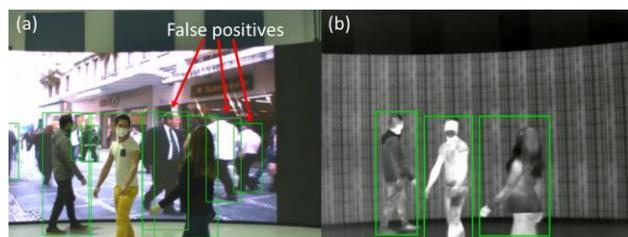

*Figure 2: Example of a scene with a fake pedestrians: (a) visible and (b) thermal.*

For example, visible cameras can be lured by images of pedestrians (e.g. a screen, advertising posters etc.). Figure 2, shows an example where the visible cameras equally classifies displays of pedestrians and real pedestrians. In this case, color information is nothing but noise.

In an aerosol-filled atmosphere (e.g. smoke, dust, fog), the contrast of the visible image is highly impacted by light scattering from ambient lighting. Thanks to the longer wavelengths, light scattering negligibly affects the thermal image. For instance, the AWARE (All Weather All Roads Enhanced vision) concluded that the LWIR band provides the best detection performances for pedestrians, cyclists, and animals in degraded weather conditions, compared to other wavebands [7].

In the case of fog or any other wet aerosol, infrared radiation is partially absorbed, which also affects the contrast, by a different physical phenomenon. Figure 3, shows an example of one of the worst weather conditions: deep winter fog that attenuates the contrast-to-noise ratio (CNR) by scattering in the visible and by absorption in the infrared. The idea is to improve the reliability of pedestrian detection by merging those two, yet degraded, orthogonal information (e.g. the pedestrian pointed with the red arrow).

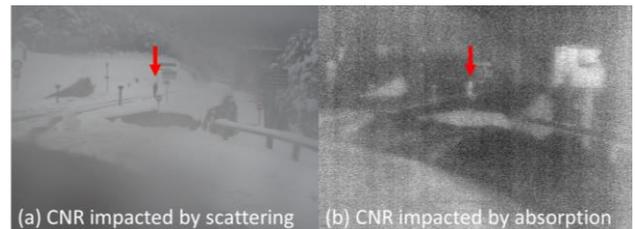

*Figure 3: Example of a snowy scene through a deep winter fog where the CNR is reduces by (a) scattering in the visible and (b) by absorption in the infrared.*

A relevant way to merge color and thermal information for computer vision is described in [8], known to as *Gated Multimodal Fusion Network*. This approach incorporates a deep neural network architecture where each modality in the color/thermal pair is processed by a CNN (fined-tuned ResNet50) dedicated to extracting features from that specific input. These features are then fused prior to the detection head. This intermediate fusion architecture enhances pedestrian detection accuracy compared to unimodal detection, early and late fusion, for both day and night.

**2.3. Field of view and resolution:** The system architect can borrow the principles from visible cameras to design the proper combination of lens and thermal image sensor: The horizontal field of view *HFOV* depends of the lens' effective focal length *EFL* and the sensor width *w*, and is estimated using the following approximation (narrow *FOV* lens with neglected distortion):

$$HFOV \approx 2\mathrm{atan}(w/2EFL) \qquad [1]$$

The instantaneous field of view (*IFOV*) depends of the pixel's pitch *p* and the *EFL*:

$$IFOV \approx \operatorname{atan}(p/EFL) \qquad [2]$$

State of the art thermal sensors have a pixel pitch of 12µm, and formats such from QVGA (320x240) to SXGA (1280x1024), sensitive to a temperature variation as low as 0.05°C. The pixel density may appear small compared to automotive grade RGB sensors. However, due to the nature of scenes based on the target to background thermal contrast, only ~20 '*pixels-on-target*' high are needed for typical deep-neural networks to detect a pedestrian with high confidence. This empirical and conservative rule will be justified in the following, and can also be found in the competitors' literature [9]. The range *d* is derived from the target height $h_0$ and the number of '*pixels-on-target*' $h_{pix}$, using the thin lens magnification formula:

$$d = \frac{h_o \cdot EFL}{h_{pix} \cdot p} \qquad [3]$$

For instance, a VGA (640x480) sensor with a wide *HFOV* of 50° is able to classify a 1.8m (5.9ft) tall pedestrian at a distance of 69m (226ft); a narrower *HFOV* of 32° can see further, at 102m (335ft).

A narrow *FOV* is well suited for long-range scenarios, with a high vehicle speed that needs a long braking distance. For this reason, we recommend a bi-focal fusion architecture (see after, on Figure 7), which involves a wide *FOV* for the visible camera, to cover low speed scenarios within the range of the headlights, and a narrow *FOV* for the thermal camera to cover long-range, high-speed scenarios.

### 3. Windshield integration

**3.1. Bumper vs. windshield (WS) integration:** Thermal cameras are generally positioned on the vehicle's bumper. This forward and lower placement offers a disadvantageous point of view (PoV) compared to a WS position.

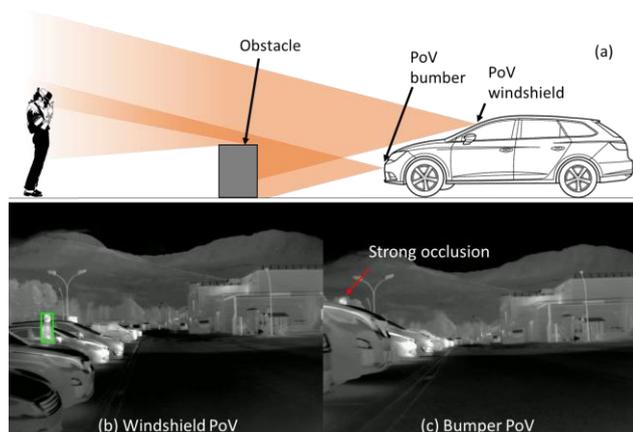

*Figure 4: Problems of occlusions with the bumper PoV. (a) Side-view schematic, (b) WS and (c) bumper PoV.*

Dangerous situations can arise where an obstacle (such as a car parked) is directly in front of the vehicle, blocking the view of a bumper-mounted camera, as illustrated in Figure 4. This position is also highly exposed to soiling. This requires a complex packaging solution, which includes a similarly intricate cleaning system, as well as a protective, defrosting IR-window that is resistant to severe abrasion.

The WS integration provides better PoV and is less exposed to soiling, thus reducing the need for an extra packaging. Additionally, fusing RGB and thermal is simplified when the two cameras are close together and share a similar PoVs, with limited parallax effects and a common mechanical reference.

**3.2. WS demonstrator:** Lynred and Saint-Gobain Sekurit partnered to develop a WS demonstrator, shown in Figure 5, compatible with thermal imaging.

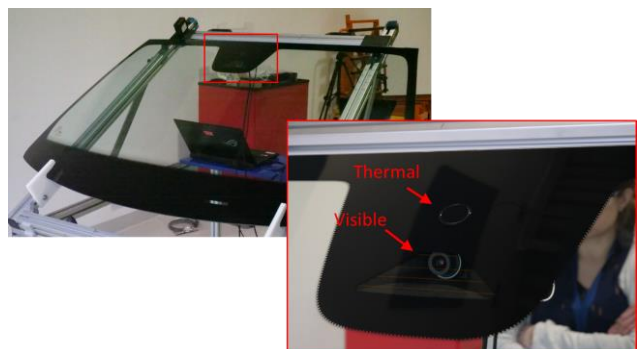

*Figure 5: WS demonstrator for thermal imaging.*

WS sensors are typically positioned in the upper central area, which is characterized by a black glazing layer on the laminated glass. The front RGB camera is placed behind a trapezoidal opening in this masked area.

Glass is not transparent in the LWIR, which is why Saint-Gobain developed a process that involves drilling the glass and inserting a 1-inch IR-transparent window, while meeting the appropriate optical, mechanical and water-sealing criteria. This insert is positioned near the visible aperture to facilitate the mechanical co-integration of both cameras and to ease the fusion of RGB/thermal images by minimizing the parallax effect, as explained earlier.

The IR-window, may be coated on the exterior face with a black hard and hydrophobic coating, adapted from proven naval warfare technologies, and resists to severe abrasion and harsh weather conditions. The transmittance the IR-window is above 70% in the 8 – 14 µm range (see, Figure 6(a)). The resulting responsivity is measured at 84.2%, accounting for the sensor's spectral sensitivity.

The exterior face has a black aspect as seen in the photography in subfigure (b), similar to the color of the masking layer of the WS, which makes the IR-window harder to notice, and more aesthetic.

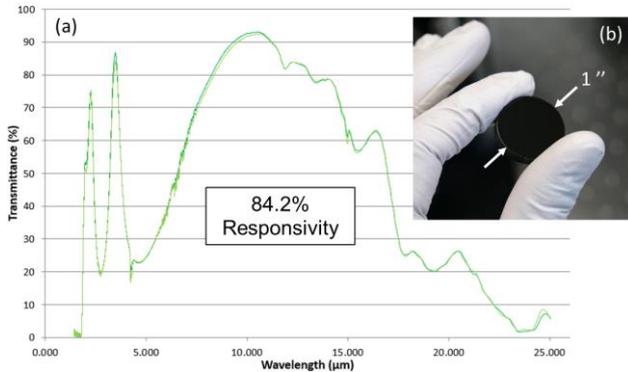

*Figure 6: (a) IR-window's transmittance spectrum, and (b) photography of the 1-inch black IR-window.*

The WS demonstrator is fitted with a thermal camera having Lynred's ATTO640 sensor (VGA, 12µm pitch) and a 32° *HFOV* objective lens (*EFL* 13.6mm f/1.0 from Ophir/MKS), adapted to the slanted aperture of the 1-inch IR-window with limited vignette effects. The color camera has the IMX273 sensor from Sony (1.57Mpix RGB, 3.45µm pitch) with a *EFL* 5mm f/2.2 objective lens from Lensation. As explained earlier, the *HFOV* of the RGB camera is wider by design than that of the thermal camera (see, Figure 7).

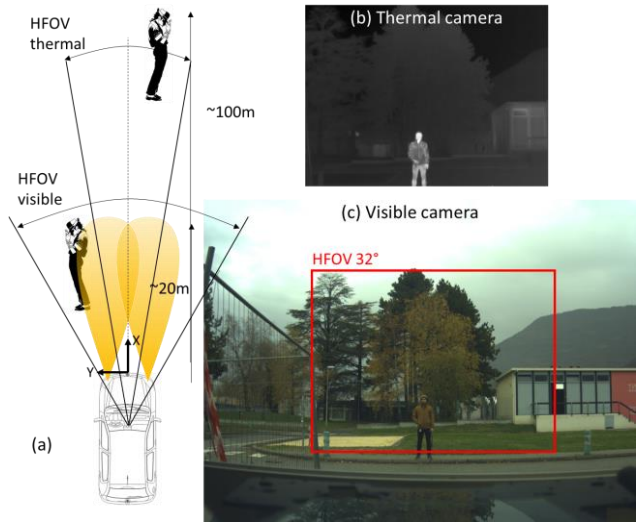

*Figure 7: (a) Top view schematic of the bi-focal fusion architecture. Images seen from the WS demonstrator: (b) thermal and (c) visible.*

**3.3. Performance degradation due to raindrops:** The presence of droplets on the WS significantly degrades the visibility of the visible front camera, which affects the operability of the AEB system during rainy weather. That is a reason why the visible front camera is covered by the wiper. Doing the same for the thermal camera (while not damaging the wiper) requires hard specifications on the flushness of the insert with respect to the glass. The question arises of whether to clean the insert with a wiper or to use an alternative/simplified cleaning method.

In Figure 8(a), we spray the WS with water. The resulting image (b) is highly distorted in the visible due to the presence of droplets, resulting in a strong performance degradation of the pedestrian classifier. The thermal image (c) remains sharp, but with a slight contrast reduction, compensated by the digital rescaling.

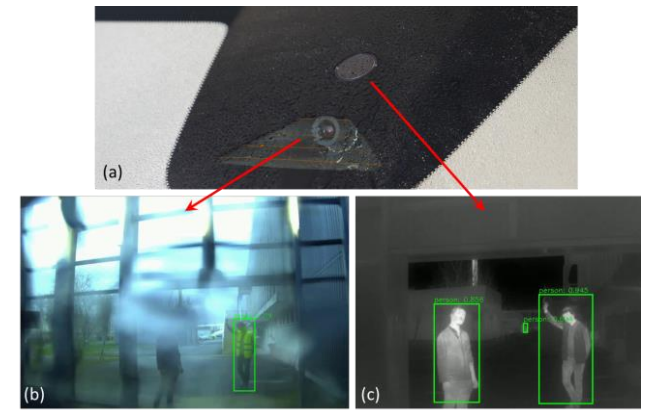

*Figure 8: Impact of droplets. (a) View of the 'wet' windshield. Images from (b) the visible camera and (c) the thermal camera, with detected pedestrians.*

We introduce simple physical properties that underline the fundamental differences in how a droplet affects the formation of an image in the visible and thermal. In Figure 9(a), we show a schematic explaining the 'lens effect', caused by a droplet. We consider water fully transparent in the visible range; it means that a droplet will act locally as an unwanted lens and distort the image captured by the visible camera, locally killing the image's sharpness.

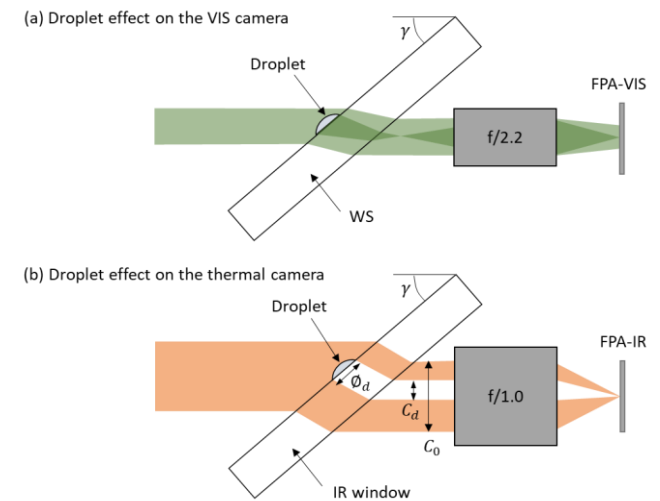

*Figure 9: Schematic of the droplet impact on (a) the visible camera, and (b) the thermal camera.*

In the thermal range, water is considered fully opaque [10]; it means that the light beam that crosses a droplet will only be masked locally, before being focused on the thermal FPA (Figure 9 (b)). In our simple model, the droplet only reduces the amount of thermal radiation that is captured by the sensor (overall contrast degradation), as a function of the droplets cross-section.

We evaluate the performance degradation of both cameras as a function of the droplet size. The demonstrator is oriented face up so that we can deposit droplets (without rolling down), a reflector is used to redirect to line of sights horizontally. We will compare the impact of droplets on the following performance indicators: classifier confidence score, spatial frequency response (SFR) and responsivity.

*Detection confidence:* For quantitative comparison, we compare visible and thermal images with a single droplet on the pedestrian. Note that the measurement of droplet size is done manually from an optical view (main source of uncertainty). In Figure 10, we show zoomed-in views of the target in both modalities, with the increasing droplet diameter $\emptyset_d$.

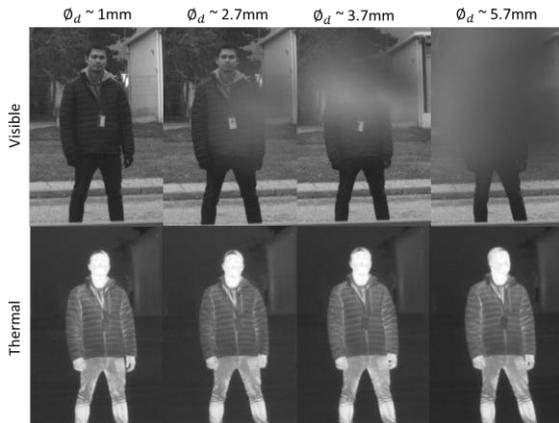

*Figure 10: Multispectral views of the target, with increasing droplet sizes.*

Inference results are shown in Figure 11.

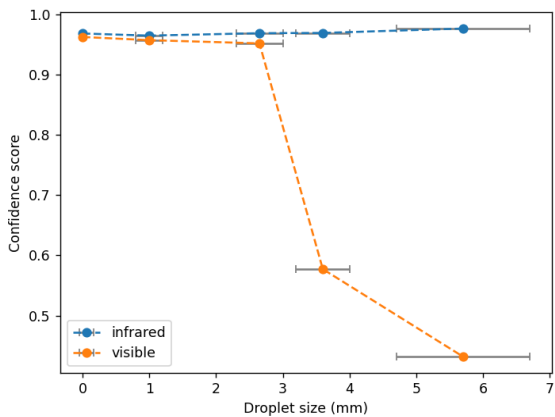

*Figure 11: Confidence score vs droplet size.*

Without droplet, the scores of both visible and thermal are very high (>95%). For droplets larger than ~3mm, the score of the visible image drop under the 80% threshold that we use to decide if the pedestrian is detected. The scores obtained with the thermal images appears unaffected by the droplet, no matter its size.

*Sharpness:* Next, we evaluate the degradation of image sharpness. This study is conducted by measuring the spatial frequency response SFR (comparable to the modulation transfer function, MTF) of each system using a slanted edge target. For the thermal, we use a 40°C blackbody (30 cm x 30 cm), covered by a slanted edge mask at ambient temperature (22°C).

We extract the SFR curves from these slanted edge images, following the procedure described in the ISO 12233 standard [11]. Note that the curves are processed from the 14-bits thermal images, and the 12-bits monochrome visible images. In Figure 12, we plot the SFR curves of both the thermal camera and the visible camera, for different droplet sizes. We show an example image of the targets, in the red frames, 'blurred' by a ~3.6mm droplet. The spatial frequency is normalized by $f_s$ the sampling frequency, in cy/mm. For monochrome sensors, such as the thermal camera $f_s = 1/p$. For sensors with a 2x2 Bayer RGB color filter we use $f_s = 1/2p$.

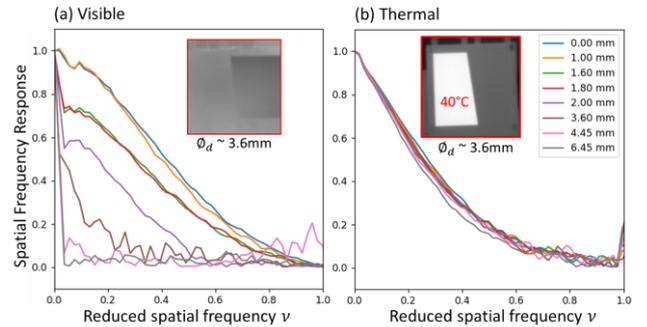

*Figure 12: SFR curves of both (a) the visible and (b) the thermal camera, for different droplet sizes.*

In Figure 13, we plot the SFR at the Nyquist frequency (half the sampling frequency, i.e. $\nu = 0.5$).

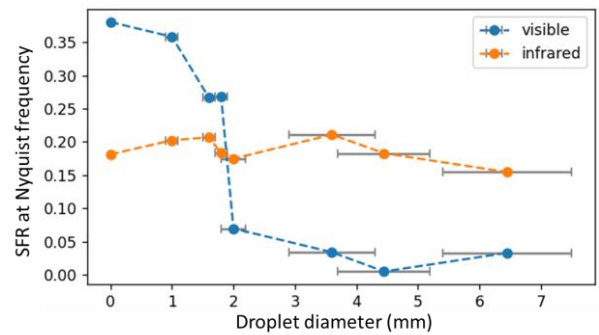

*Figure 13: SFR at Nyquist frequency vs. droplet size.*

Without droplets, the SFR is better with the visible, notably because the thermal camera is limited by diffraction (longer wavelengths) and because the target is closer than the lens' hyperfocal distance. When the droplet size increase, the visible's SFR has a strong degradation, for a droplet size of about 2 mm. The thermal's SFR shows no signs of degradation at these droplet sizes.

*Responsivity:* The responsivity is a metric that relates the sensitivity of the thermal camera output to a variation of the scene temperature. In this section, we evaluate the impact of the droplets on the camera's responsivity $R(\emptyset_p)$. This responsivity directly affects the contrast-to-noise ratio (CNR), which is a key performance metric for detecting a pedestrian.

$$CNR(\emptyset_p) \approx \frac{\Delta T \cdot R(\emptyset_p)}{\sigma} \quad [4]$$

Where, $\Delta T$ is the pedestrian to background thermal contrast and $\sigma$ is the camera's temporal noise.

We measure the signal contrast between a 40°C blackbody and a 25°C reference blackbody. The variation of the camera responsivity as a function of the droplet's diameter is shown in Figure 14.

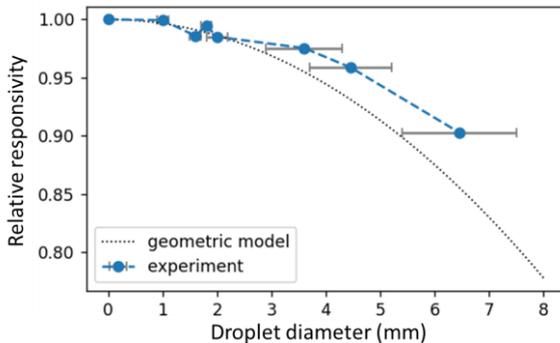

*Figure 14: Relative responsivity vs. the droplet diameter.*

The relative responsivity is compared with a simplistic geometric occlusion model (see, Figure 9(b)), based on the ratio of the droplet's cross-section $C_d$ and the clear optical cross-section $C_0$, which expression is given in eq. [4].

$$T(\emptyset_d) \approx 1 - \frac{C_d}{C_0} = 1 - \frac{\emptyset_d^2 \sin\gamma}{(EFL/f_\#)^2} \quad [5]$$

We see a slight drop in responsivity that follows the same tendencies as our geometric transmittance (uncertainty in our model is expected). The degradation of responsivity is measured at 90% responsivity for a large droplet of ~6mm. This is more than acceptable given that the pedestrian to background thermal contrast is good, especially during wet/cold weather.

*Conclusion on cleaning system:* This study has shown that while raindrops have a critical impact on RGB cameras performances, their effect on thermal cameras is very limited. Consequently, a simpler solution for cleaning can be used, including solutions without wipers (non-contact cleaning, coating etc.). Even without cleaning, system reliability (ability to classify) maintains high performances, which ensures operational safety and availability of the function whatever the rain conditions and wiping/cleaning frequency.

## 4. Thermal's camera minimal required performances for AEB

Thermal imaging is already well implemented in ADAS, as a luxury option that allows the driver to have night vision displayed on a screen. To meet the new safety standards imposed by the NHSTA, it is important to understand the minimum performance requirements for AEB functions, such as the pedestrian detection range. The focus is no longer on image quality perceived by the driver, based on psycho-cognitive criteria, but rather on the performance of the machine perception system, in other words, on AI performance. Lowering the SWaP relies on the following key parameters: image resolution and pixel pitch. The sensor size reduction increases the number of sensors produced in parallel on a single wafer, a well-known concept in the microelectronics industry. A smaller sensor also reduces the size of imaging lenses, with certain components benefiting from parallel manufacturing steps as well.

### 4.1. Sensor's resolution: VGA or QVGA?

The VGA image format is the current state of the art for the luxury night-vision option. Let's see if this format can be reduced. This section presents pedestrian range tests conducted in collaboration with the Europe's JRC (Joint Research Laboratory), more information are found in reference [12]. We evaluate the classifier's performance for two thermal camera configurations, with Lynred's sensors and Umicore's objective lenses, characteristics are given in Table 1:

| Format    | QVGA (320x240) | VGA (640x480) |
|-----------|----------------|---------------|
| Pitch (µm)| 12 µm                          ||
| EFL (mm)  | 6.2 mm         | 14 mm         |
| f/#       | f/1.0                          ||
| HFOV (°)  | 36°            | 31°           |

*Table 1: Camera configurations with two resolutions.*

Different targets (a 1.65m woman, 1.75m man, cyclist, 1.7m man with a german shepherd) where recorded at different distances, in two environments (rural and urban). The trials were made during the day (ambient luminosity >400lux), at dusk (400-5lux) and at night (<5lux).

In Figure 15, we plot the averaged confidence score for all targets and conditions, as a function of the target distance. We report no significant differences in confidence depending on the lighting conditions (day, dusk or night) for both thermal cameras.

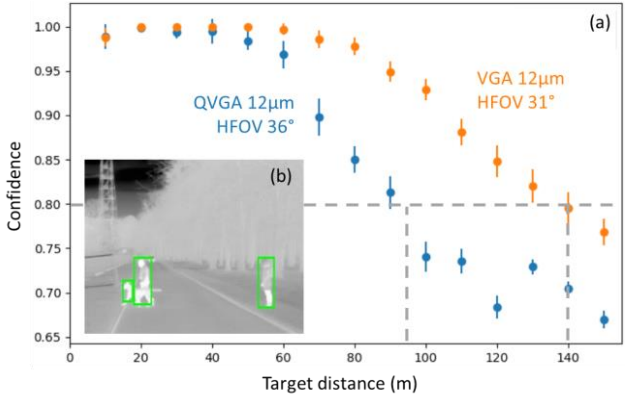

*Figure 15: (a) Confidence score vs. distance: QVGA/VGA. (b) Example of inferred image.*

By setting a 80% confidence threshold, we find a detection range of 90m (295ft) for the QVGA configuration, and 140m (460ft) for the VGA. The detection threshold, which impacts the detection range, must be chosen as a compromise between the miss rate and the false detection rate. Using eq. [3], with a mean target height of 1.7m, we find a minimal height of 10px for the QVGA configuration and 15px for the VGA configuration. Note that the QVGA configuration outperforms our prior estimations. Both configurations are within our empirical rule of '*20 pixels-on-target*'.

**4.2. Scaling down the pitch from 12µm to 8.5µm**

The 12µm pixel pitch represents the current state of the art, and the next generation of small pixel pitch at 8.5µm is well advanced in development at Lynred [13], with both a VGA and a QVGA prototypes, that already provide high quality images (see, Figure 16).

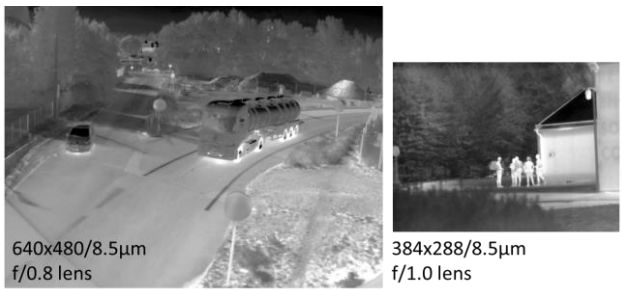

*Figure 16: Images examples of 8.5µm pitch Lynred's prototypes. (right) VGA with a f/0.8 lens, (b) QVGA with a f/1.0 lens.*

One might assume that compensating for the reduction in pixel surface area requires increasing the aperture size of the imaging lens in order to maintain equivalent performances, notably in terms of diffraction (pixel smaller than the mean wavelength). An equivalent configuration would pair a 12µm pixel with an f/1.0 lens, and an 8.5µm pixel with an f/0.7 lens, but increasing the lens' aperture comes with a number of negative effects [14]. Scaling down the lens design by 70%, while keeping a lens aperture of f/1.0 might still be a relevant choice, when considering a task specific performance metric such as the pedestrian detection range.

In this section, we evaluate the impact of reducing the pixel pitch from 12µm to 8.5µm in terms of range performance, using the same f/1.0 aperture lens (see, Table 2), with similar FOVs.

| Pitch (µm)   | 12µm            | 8.5µm  |
|--------------|-----------------|--------|
| Image format | VGA (640x480)   |        |
| EFL          | 14mm            | 8.8mm  |
| f/#          | f/1.0           |        |
| HFOV (°)     | 31°             | 34°    |

*Table 2: Camera configurations: 12µm vs 8.5µm pitch.*

We compare view of a 1.8m tall pedestrian with both cameras, while increasing the distance (Figure 17(b)). It should be noted that the test conditions were those of a hot summer day, which represents a disadvantageous situation in terms of thermal contrast (pedestrian's temperature close to the ambient temperature), and not the most relevant use-case for AEB.

As expected, the 8.5µm images appears blurry compared with the 12µm images, due to diffraction. Inference results are given in Figure 17(a), showing no performance degradation of the pitch reduction, this validates the f/1.0 lens choice.

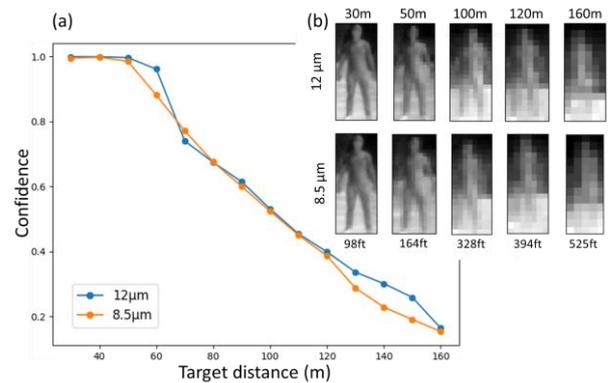

*Figure 17: (a) Confidence vs. distance: pitch 12µm/8.5µm. (b) Zoomed view of the pedestrian at different distances with both cameras.*

**4.3. A low-SWaP thermal camera for AEB**

Minimal performances requirements and resulting specifications are detailed in [15], and summarized in Table 3. The thermal camera has to be designed to cover high-speed scenarios up to 60km/h (37.5mph), with a required detection range >46m (>151ft), that corresponds to the car's stopping distance (0.5g deceleration, AEB reaction time: 1s).

| Image format | QVGA 16:9 (320x180) |
|---|---|
| Pixel pitch | 8.5µm |
| HFOV x VFOV | 37° x 20° |
| EFL, f/# | 4.2mm, f/1.0 |

*Table 3: Configuration of the minimal thermal AEB camera*

If we are using a small format sensor with an 8.5µm pixel pitch, it allows us to minimize the size of the optics. Lens elements smaller than 5mm can be molded from chalcogenide glass and coated wafer level (WLO, wafer level optics) [16], [17]. Example of a diced WLO lens element is shown in Figure 18(a).

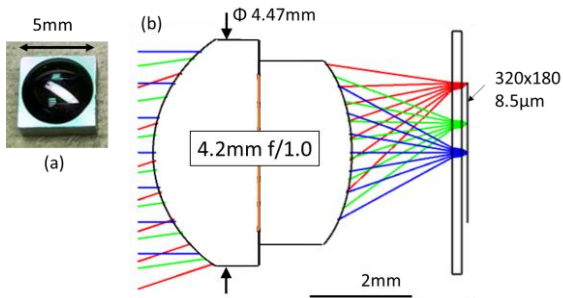

*Figure 18: (a) Example of a chalcogenide WLO element. (b) 4.2mm f/1.0 lens design (censored) with a QVGA 16:9 8.5µm pitch sensor.*

As for a design example, we can simply scale down an existing f/1.0 lens to match the new smaller sensor, while keeping the same aperture (Umicore's 14mm f/1.0 chalcogenide lens scaled down by 30%). The entrance pupil diameter is kept below 5mm, which makes it compatible WLO manufacturing technology. All optical simulations are made with the Zemax Optic Studio software (Figure 18(b)).

In Figure 19, we show an image simulation of the lens with the design example, with an image quality in-line with what's expected from QVGA.

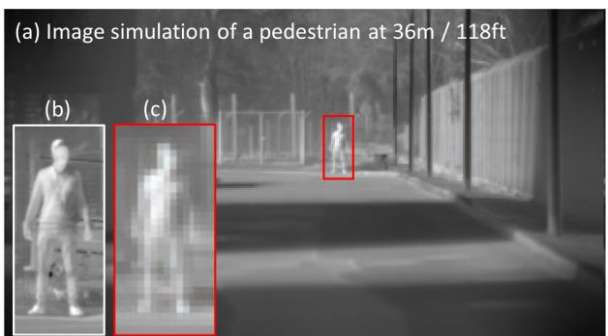

*Figure 19: (a) Image simulation of a pedestrian at 36m. Zoomed reference, assumed perfect (b) vs. simulated (c) images.*

This design is compatible with a WS integration. For instance, it can be fitted behind a 20mm diameter IR-window (simplified), with a WS inclination angle of 28.1°, without vignette effects (see, Figure 20).

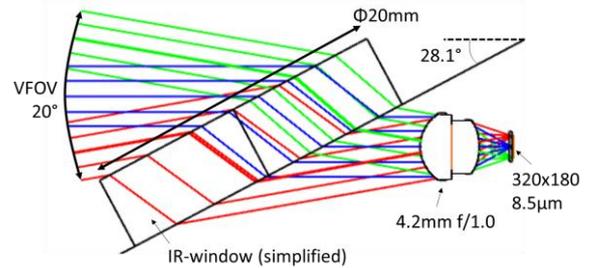

*Figure 20: WS compatibility of the proposed camera design.*

### 5. Conclusion

This document highlights the value of thermal imaging for nighttime AEB, in response to the latest regulations imposed by the NHTSA.

Thermal imaging, when fused with visible imaging, enhances the reliability of AEB systems in both day and night, and in low visibility conditions (rain, fog, glaring etc.). We demonstrate the integration of a thermal camera into the windshield (WS), in a position close to the front visible camera. Compared to the conventional bumper integration, windshield integration reduces soiling exposure, simplifying cleaning systems and packaging. This higher position provides a point of view less prone to occlusions and closer to the visible camera, thereby minimizing parallax effects. We demonstrate that the thermal camera maintains its performances even with large raindrops on the WS: this ensures the system's reliability and availability even with a simplified cleaning method.

Through field tests of AI performance in pedestrian detection range, we evaluate the minimum performance requirements for a thermal camera that meets NHTSA standards for AEB. These minimum performance levels, combined with the latest technological advancements in the manufacturing of 8.5µm pixel pitch sensors and wafer-level molded chalcogenide optics, allow us to propose a compact camera design compatible with small IR-window and windshield integration.

### 6. Acknowledgement

The authors are thankful to the JRC, in Ispra (Italy) for their collaborations in the trials in section 4.1; to Christophe Kleo and Laurent Silvestrini from SGR Compiegne for their support on the development of the WS demonstrator; to Nicolas Mielec from SGR Paris for ensuring the continuity of the development work on the WS demonstrator.

# 7. References

[1] European Road Safety Observatory, "Facts and Figures – Pedestrians - 2021," Brussels, 2021.
[2] NHTSA, "Overview of Motor Vehicle Traffic Crashes in 2021," DOT HS 813 435.
[3] NHTSA, "Federal Motor Vehicle Safety Standards: Automatic emergency Braking Systems for Light Vehicles," RIN 2127-AM37, Apr. 2024. [Online]. Available: https://www.nhtsa.gov/sites/nhtsa.gov/files/2023-05/AEB-NPRM-Web-Version-05-31-2023.pdf
[4] K. He, X. Zhang, S. Ren, and J. Sun, "Deep Residual Learning for Image Recognition," presented at the IEEE Conference on Computer Vision and Pattern Recognition (CVPR), Las Vegas, NV, USA: IEEE, Jun. 2016. doi: 10.1109/CVPR.2016.90.
[5] C. Karam, J. Matias, X. Brenière, and J. Chanussot, "Optimizing the image correction pipeline for pedestrian detection in the thermal-infrared domain," Jul. 2024, doi: https://doi.org/10.48550/arXiv.2407.04484.
[6] FLIR and VSI Labs, "Fused AEB with thermal can save lives: A test summary for testing agencies, automotive manufacturers, and tier 1 suppliers." [Online]. Available: https://www.flir.com/globalassets/industrial/oem/adas/flir-thermal-aeb-white-paper---final-v1.pdf
[7] N. Pinchon et al., "All-weather vision for automotive safety: which spectral band?," presented at the AMAA 2018, Advanced Microsystems for Automotive Applications, Sep. 2018. doi: 10.1007/978-3-319-99762-9_1.
[8] A. Chadli, E. Balit, P. Arquier, G. Delubac, E. Bercier, and X. Brenière, "Gated Multimodal Fusion for Visible-Thermal Pedestrian Detection," Sep. 2020, [Online]. Available: https://neovision.fr/wp-content/uploads/2022/01/sia_vision_paper2020_final-2_compressed.pdf
[9] FLIR, "Why ADAS and autonomous Vehicles need thermal infrared cameras," Dec. 2018. [Online]. Available: https://www.flir.com/globalassets/email-assets/pdf/flir_thermal_for_adas_and_av.pdf
[10] "Optical Absorption of Water Compendium." [Online]. Available: https://omlc.org/spectra/water/abs/index.html
[11] *ISO12233 : Photography - Electronic Still Picture Imaging - Resolution and Spatial Frequency Response*, 2017.
[12] R. Donà, K. Mattas, S. Vass, G. Delubac, J. Matias, and S. Tinnes, "On the Assessment of Thermal Cameras and their Safety Implications for Pedestrian Protection: a mixed Empirical and Simulation-based characterization," presented at the TRB Annual Meeting 2024, 2024.
[13] S. Cortial, V. Gorge, and C. Pautet, "High performances 8.5 microns pitch microbolometers focal plane arrays," presented at the OPTRO 2024, 11th International Symposium on Optronics in Defense & Security, Bordeaux, France: 3AF, Jan. 2023.
[14] R. Proux and J. W. Franks, "Lens requirements for sub-10μm pixel pitch uncooled microbolometers," in *Proceedings Volume 12533, Infrared Imaging Systems: Design, Analysis, Modeling, and Testing XXXIV*, Orlando, Florida, United States, Jun. 2023. doi: https://doi.org/10.1117/12.2665473.
[15] S. Tinnes, Q. Noir, G. Delubac, J. Matias, R. Donà, and B. Ciuffo, "Automatic Emergency Braking: How can affordable thermal camera improve reliability and extend use cases to nighttime conditions," in *Proceedings Volume 13046, Infrared Technology and Applications L; 130460Y*, National Harbor, Maryland, United States, Apr. 2024. doi: https://doi.org/10.1117/12.3013738.
[16] J. Franks, "Molded, wafer level optics for long wave infra-red applications," in *Proceedings Volume 9822, Advanced Optics for Defense Applications: UV through LWIR*, Baltimore, MD, United States, 2016. doi: https://doi.org/10.1117/12.2223872.
[17] G. Druart et al., "Study of infrared hybrid Chalcogenide Silicon lenses compatible with wafer-level manufacturing process for automotive application," presented at the SPIE Optical Systems Design 2021, United States, Sep. 2021. doi: 10.1117/12.2596888.

# 8. Glossary

| | |
|---|---|
| *ADAS* | Advanced Driver Assistance System |
| *AEB* | Automatic Emergency Braking |
| *AI* | Artificial Intelligence |
| *CNN* | Convolutional Neural Network |
| *CNR* | Contrast-to-Noise Ratio |
| *EFL* | Effective Focal Length |
| *FOV* | Field of View |
| *FPA* | Focal Plane Array |
| *HFOV, VFOV, IFOV* | Horizontal, Vertical, Instantaneous FOV |
| *IR* | InfraRed |
| *LWIR* | Long-Wave IR |
| *MTF* | Modulation Transfer Function |
| *NHSTA* | National Highway Traffic Safety Administration |
| *PoV* | Point-of-View |
| *QVGA* | Quarter-VGA |
| *RGB* | Red-Green-Blue |
| *SFR* | Spatial Frequency Response |
| *SWaP* | Size Weight and Power |
| *VGA* | Video Graphics Array |
| *WS* | Windshield |